\documentclass[preprint,12pt]{elsarticle}

\usepackage{amssymb}

\usepackage{amsmath,amsxtra}
\usepackage{epsfig}
\usepackage{epic}
\usepackage[dvipsnames]{xcolor}
\definecolor{bordo}{rgb}{0.8,0.3,0.3}
\definecolor{azul}{rgb}{0.1,0.2,0.7}
\usepackage[english]{babel}
\usepackage{color}
\usepackage{verbatim}
\definecolor{azul}{rgb}{0.1,0.2,0.6} 
\definecolor{verde}{rgb}{0.1,0.5,0.3}
\definecolor{bordo}{rgb}{0.7,0.3,0.3}
\usepackage{lipsum}
\usepackage[autostyle=false, style=english]{csquotes}
\MakeOuterQuote{"}
\usepackage{mathrsfs}
\usepackage[scr=dutchcal,calscaled=1]{mathalfa}
\usepackage[section]{placeins}
\usepackage{float}
\usepackage[hyphens]{url}
\journal{Arxiv}

\begin{document}

\begin{frontmatter}




\title{{An algorithm to represent inbreeding trees}}


\author[label1]{C. Jarne}
\address[label1]{Universidad Nacional de Quilmes, Departamento de Ciencia y Tecnolog\'ia CONICET}
\ead{cecilia.jarne@unq.edu.ar}

\author[label2]{F. A. G\'omez Albarrac\'in}
\address[label2]{Instituto de F\'isica de L\'iquidos y Sistemas Biol\'ogicos (IFLYSIB), UNLP-CONICET, La Plata, Argentina and Departamento de F\'isica, Facultad de Ciencias Exactas, Universidad Nacional de La Plata; Departamento de Ciencias B\'asicas, Facultad de Ingenier\'ia, UNLP}

\author[label3]{M. Caruso}
\address[label3]{Departamento de Estad\'istica e Investigaci\'on Operativa, Universidad de Granada, Campus de Fuentenueva, Espa\~na}

\begin{abstract}

Recent work has proven the existence of extreme inbreeding in a European ancestry sample taken from the contemporary UK population \cite{nature_01}. This result brings our attention again to a math problem related to inbreeding family trees and diversity. Groups with a finite number of individuals could give a variety of genetic relationships.  { In previous works \cite{PhysRevE.92.052132, PhysRevE.90.022125, JARNE20191}, we have addressed the issue of building inbreeding trees for biparental reproduction using Markovian models. Here, we extend these studies by presenting an algorithm to generate and represent inbreeding trees with no overlapping generations. We explicitly assume a two-gender reproductory scheme, and we pay particular attention to the links between nodes. We show that even for a simple case with a relatively small number of nodes in the tree, there are a large number of possible ways to rearrange the links between generations. We present an open-source python code to generate the tree graph, the adjacency matrix, and the histogram of the links for each different tree representation. We show how this mapping reflects the difference between tree realizations, and how valuable information may be extracted upon inspection of these matrices. The algorithm includes a feature to average several tree realizations, obtain the connectivity distribution, and calculate the average and mean value. We used this feature to compare trees with a different number of generations and nodes. The code presented here, available in Git-Hub, may be easily modified to be applied to other areas of interest involving connections between individuals, extend the study to add more characteristics of the different nodes, etc.

}

\end{abstract}




\end{frontmatter}


\section{Introduction}\label{intro}

Several studies were conducted regarding the inbreeding strategy of different animals. They have reported from inbreeding tolerance in different species \cite{11-cell,16-cell} to inbreeding preference for some of them \cite{17-cell}. Recent studies have also found evidence of regular incest behavior in wild mammals \cite{otro}. Even more interesting is the fact that the existence of extreme inbreeding in humans was also detected when considering a European ancestry sample taken from the contemporary UK population \cite{nature_01}. 

{In our previous studies \cite{PhysRevE.92.052132, PhysRevE.90.022125, JARNE20191}, we simulated random trees of ancestors considering inbreeding along the tree. To build the tree, generations were added in a markovian way.} The number of ancestors, for a given generation, was a random quantity limited to a maximum value given by the case where all ancestors were different. In that approach, the only constraint considered in the number of ancestors was random blood relationships between individuals of the same generation (there was no distinction between the nodes in each generation). Other restrictions in the number of ancestors related to culture, in the human case, ethological in the animal case, or regarding isolation of populations, were not taken into account. {Most importantly, the model presented a  random mating in nonoverlapping generations with negligible mutation and selection. These two assumptions are common to develop population genetic models; in particular, these are present in the Hardy-Weinberg principle \cite{Bacaer2011}.
}

In the present work, we deepen our previous studies by presenting a new algorithm. {Firstly, we take into account the gender of the individuals, supposing a binary female/male reproduction scheme (which can be changed by the user), ensuring that each individual has one female and one male direct ancestor. Secondly, we explore the different inbreeding cases considering the different links between consecutive generations. We show with a simple example that this number can be quite large. The algorithm produces different inbreeding trees, following a given set of rules, and also generates a mapping of each tree realization to an adjacency matrix and a link histogram. We show how these matrices and histograms reflect different characteristics of the trees and are very useful tools to represent the links between generations. Moreover, the algorithm includes a feature to average tree realizations, obtain the average output link distribution, its histogram, mean value, and standard deviation. } The code is available online, and it can be easily used and modified.
 
{The manuscript is organized as follows. In Section \ref{one_tree}, we discuss the model and define the rules to obtain tree realizations or representations.}  We then describe the algorithm {to build tree representations and its} computational implementation. The adjacency matrix representation is discussed in Section \ref{properties}. In that section, we profit from the flexibility of the code presented in this work to generate different cases of trees and comment on their different characteristics in the adjacency matrix representation and the construction of link histograms. Finally, we close in Section \ref{conclu} with conclusions and discuss future work.

\section{ { {Model and algorithm}}} \label{one_tree}

{
In this work, we are concerned with building trees of ancestors. Considering a biparental reproductive scheme, the full binary tree model (where individuals do not have common ancestors) gives an increasing number of ancestors, which is clearly not realistic. Some degree of connection between nodes (inbreeding) is necessary. Here we present an algorithm to build different trees of ancestors, under certain assumptions and rules, which we describe below.

Our model assumes three conditions:
\begin{enumerate}
 \item[Condition 1:] { There is no overlapping between generations \cite{Bacaer2011}. In this way, we maintain the concept of ``generation''}
 \item[Condition 2:]  { Since we are interested in representing trees of ancestors, in each tree, every node is linked to both the preceding and the next generation (except of course those in the extremes of the tree), i. e.: there are no ``lose'' nodes. }
 \item[Condition 3:]  { We assume a binary animal reproductive scheme, where each individual has two ancestors.  We distinguish between two reproductory genders, male and female, and for each individual in each generation, a female and a male direct ancestor are required.}
\end{enumerate}
}

As we discuss later, these conditions can be changed in the algorithm to take into account different situations.

\begin{figure}[htb!]
\begin{center}
\includegraphics[trim={0 2.5cm 0 1.5cm},clip,width=1.15\textwidth]{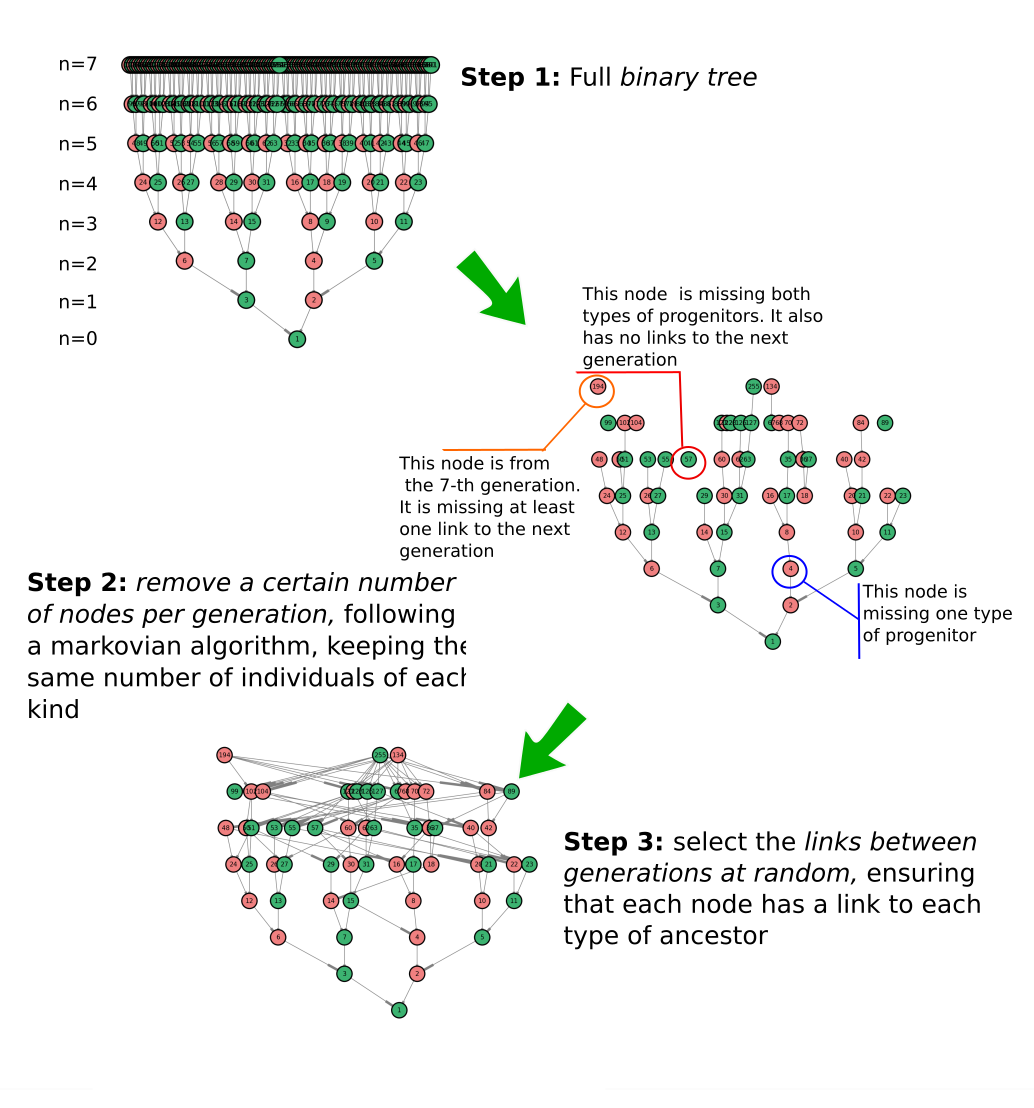}

\caption{ Sketch of the process to build one possible ancestors' tree with inbreeding. At the top, step 1, we show the full binary tree. Each ancestor is represented with a node. Pink nodes labeled with even numbers represent female animals. Green nodes with odd numbers represent male animals. The tree starts from the present generation at the bottom, labeled $n=0$, to the seventh generation at the top ($n=7$). The nodes are numbered from the bottom to the top. In the middle, step 2, there is an example of a tree where some of the ancestors (nodes) have been removed randomly (following certain rules, see text for details) from the previous full binary tree.  Nodes missing different types of links are indicated. In Step 3, at the bottom of the figure, we show one possible tree realization: one way to reconnect all the nodes, following the assumptions of the model. }
\label{fig_00}
\end{center}
\end{figure}

Now, we proceed to construct one possible tree {realization}. { We call a  ``realization'' (or representation) of the tree to one of the possible connections of the links from a given number of ancestors in the tree and generations.} To do this, we start from a full binary tree. { We illustrated a full binary 7-generation tree  in Figure \ref{fig_00} (top). We included in the figure the numbering of the generations: $n=0$ is the individual whose tree of ancestors we are building, $n=7$ is the ``oldest'' generation of ancestors. Clearly, at the $n$-th generation, there are $2^n$ ancestors. As seen in the figure},   to uniquely identify each individual, we assign each of them a number and distinguish {both types of ancestors} (males and females) with odd and even numbers, respectively.  
{ The nodes are numbered considering the label 1 for the first individual of the ground generation. Starting from there, we proceed incrementally by labeling all the ancestors of the binary tree in this way: for each node with label $x$, the female direct ancestor is labeled $2x$ and the male direct ancestor $2x+1$. The number assigned to each ancestor is used later in the algorithm as a tag to randomly remove an ancestor in each generation of the tree.}

Then, we remove in each generation { a given number of nodes and links (we will discuss the specific way that nodes are removed in this work in the description of the algorithm)}. In Figure \ref{fig_00} (middle) we show the resulting tree, obtained after subtracting a certain number of nodes from the binary tree. Due to this process, some individuals will lose the links, which represent the ``progenitor-offspring'' relationship between generations. In Figure \ref{fig_00} (middle) we indicate different types of missing links in the nodes. The goal now is to distribute all the missing links between generations among the current nodes. These nodes will be the progenitors of those members who lost a link in the tree when removing individuals. These new links represent inbreeding relationships since they will not follow the full binary tree. { An example of a possible inbreeding tree obtained from that shown in Figure \ref{fig_00} (middle) is illustrated at the bottom of Figure \ref{fig_00}. }

{ The procedure to build a realization of a tree, illustrated in Figure \ref{fig_00}, can be summarized as follows:

\begin{enumerate}
 \item[Step 1:]{ We start from a full binary tree with the number of generations we are interested in}
 \item[Step 2:]{ We remove nodes and links in each generation}
 \item[Step 3:]{ We reconnect the remaining nodes, in such a way that the tree conditions of the model are fulfilled (there is no overlapping or ``jumping'' between generations, there are no ``lose'' nodes, and each node has two direct ancestors, one male, and one female)}
\end{enumerate}

}

{ Clearly, there are many possible ways to fulfill these steps and obtain a tree representation, even following the three conditions from our model. Firstly, there are many ways to remove nodes and links from a full binary tree. Secondly, once removed, there are also many ways to reconnect the nodes, keeping in mind the conditions.  Let us do a quick inspection of the ``incomplete tree'' from Figure \ref{fig_00} (middle) to show this. For example, the node labeled $4$ at the $n=2$ generation is missing a male direct ancestor, and there are three possible candidates in $n=3$. In $n=3$, node $14$ is missing a link to a direct female ancestor. Even if there are 7 ancestors at $n=4$, in this particular case, the only possible link is to node $18$ since $18$ is the only female node at this generation missing a link to $n=3$ (Condition 2).  At $n=5$, node $57$ is the only male node missing a link to $n=4$. There are 7 candidates at $n=4$ that are missing a male direct ancestor: nodes $24,30,16,20,21,22,23$. Once a link is chosen, for each of the remaining six nodes at $n=4$, there are 7 possible male direct ancestors at $n=5$. Since it is not necessary for them to have different male ancestors, there are $7^6$ possible ways to link them. Repeating this analysis at every generation shows that the number of possible trees is significantly large. 

Here, in the next subsection, we present in detail our algorithm to generate different tree representations and define the rules used to construct them.}

\subsection{ {Algorithm description}}
{We now proceed describing in detail each step of the  algorithm to construct one realization for the tree of ancestors with inbreeding, which is represented in Figure \ref{fig_00}}

\begin{enumerate}

 \item We start from a full binary tree with a given number of generations. { In terms of code, it means a list of nodes and links. In this tree, each member has an associated tag (a number), as explained above. Here a binary tree could be trivially built using any graph code generator.}
 \item From the full binary tree, we remove { individuals (nodes) in each generation. We follow the ideas from our previous work  \cite{JARNE20191}: we remove nodes from generations $n\geq 3$, using a markovian algorithm, in such a way that there is a generation that holds the largest number of nodes, and more nodes are removed in the ``older'' generations (larger $n$). In the present work, an important feature is Condition 3 from our model: the fact that there are two genders and one direct ancestor of each kind is needed. This means that } we have to  {decide}  how many labels or tags representing each gender we will extract. Here { we introduce} the constraint that the number of both types of individuals remains the same, or close if the number is odd. {Different reproductive behaviors, such as keeping a few nodes of one gender and multiple nodes of the other gender per generation, or other constraints (for example, that one type of gender may only have one offspring), could be considered in this step.}
 \item  { Now,} as some nodes were extracted, the links involving them disappear, as shown in Figure \ref{fig_00} (middle). { We do not remove additional links. To construct a tree realization,}  we must also choose a {set of rules} to distribute the missing links. 
 
 \begin{itemize}
 \item[(A)] { First, we check that each node has at least an output link (i.e.: if it is indeed an ancestor). We go over the nodes in each generation and check if there are output links. For a node that has no output links to the next generation, we check its gender label. We look at the generation of the descendants which nodes are missing links to a direct descendant with this label, and we randomly assigned to the ancestor an output link to one of these nodes. We only do this for elements that are missing links to the generation of the descendants.
 
 \item[(B)] Then, we} ensure that each node has a direct ancestor of each gender as follows: we go over each of the generations of the tree except the last one. In each generation, we analyze each node one by one. We observe how many links arrive at the node and whether these links come from a node with a female or male gender label. With this information, we decide what to do:
 
\begin{enumerate}

\item If there is no link arriving at the node from the previous generation, we randomly choose a node of the previous one to be its ancestor.

\item If there is a link from the node of the previous generation, we check the label from this ancestor and randomly choose an individual from that generation among those with the opposite label.

\item If the node considered already has the 2 links, we do not add anything.

\end{enumerate}

When we finish, we move to the next generation and repeat the process, thus going through the entire tree.
\end{itemize}

In this way, we can obtain one representation, like the one shown at the bottom of Figure \ref{fig_00}. 

\end{enumerate}

As we have mentioned, there are multiple ways to distribute the links between generations among the existing nodes. In this way,  trees with very different kinship relationships can be created,  keeping the same amount of ancestors. 

{ We emphasize that }, to build the representation the program needs as input the number of ancestors per generation. This number may be obtained using the code presented in \cite{JARNE20191}, { as was done in this work,} or fixed at will, { in order to explore other reproductory behaviors or compare specific structures in trees, etc.}. With this information, the program generates one {realization}. The output of the program is: i) one of all possible { trees} ii) its corresponding adjacency matrix, and node distributions \cite{normanbiggs1994}, which we present and discuss in the next section. { We will also show the results of averaging tree representations, a feature that is included in the program.}

\begin{figure}[htb!]
\begin{center}
\includegraphics[width=15cm]{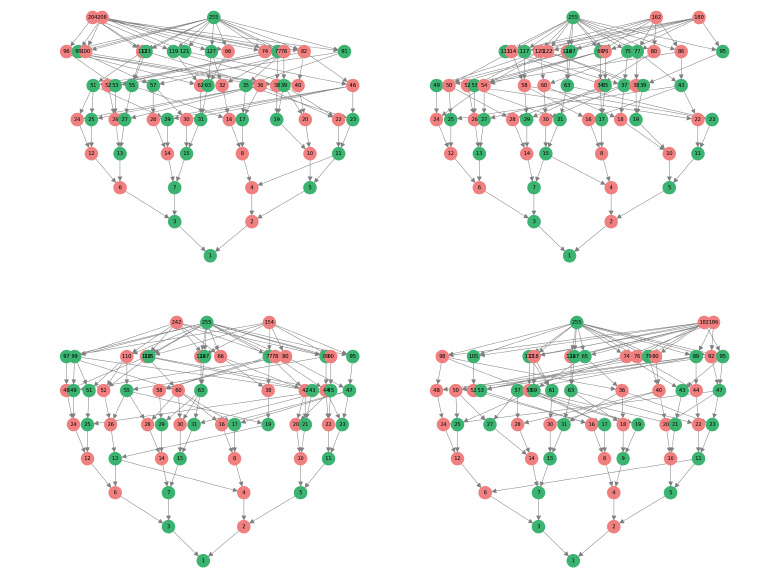}
\caption{Four possible different realizations of 7-generation trees for the same number of ancestors in each generation.}
\label{fig_02}
\end{center}
\end{figure}

{ The representation does not depend only on the number of ancestors in each generation and gender labels, but} it also depends on the seed we put in the random number generator. Different seed values give rise to different representations like those observed in Figure \ref{fig_02}. { These representations were constructed fixing the number of nodes at each generation, in order to compare them.}

\subsection{{Algorithm implementation}}

{The algorithm was developed and implemented   using Python \cite{Rossum:1995:PRM:869369} and Networkx \cite{team2014networkx}. It is available in an open-source repository at Github. It may be used and modified freely. Plots and analysis were also performed using Python and Matplotlib \cite{Hunter:2007}. All Figures in the paper can be reproduced using the available code.}

\section{{ Visualization of tree representations}} \label{properties}

{In this section, we explore how the different realizations of the inbreeding trees could be visualized to extract relevant information.  To do that}, let us consider the adjacency matrix for a simple graph with vertex set $\mathbf{V}$. It is a square $|\mathbf{V}|\times|\mathbf{V}|$ matrix $\pmb{\mathcal{M}}$ where the element $\mathcal{M}_{ij}$ is 1 when there is a edge from the $i-$vertex to the $j-$vertex, and 0 when there is no edge \cite{normanbiggs1994}, where $|\mathbf{V}|$ is the number of elements of $\mathbf{V}$. We use this definition with the matrix of each directed graph representing the tree in such a way that columns represent the relationships between generations. Each column always has two non-null elements representing each progenitor. Each row indicated how many nodes from the next generation are linked to a certain node {of the present generation (offspring) of each progenitor. In such a way,} the vertical axis represents the nodes as ancestors and the horizontal axis as offspring.

\begin{figure}[htb!]
\begin{center}
\includegraphics[width=0.9\textwidth]{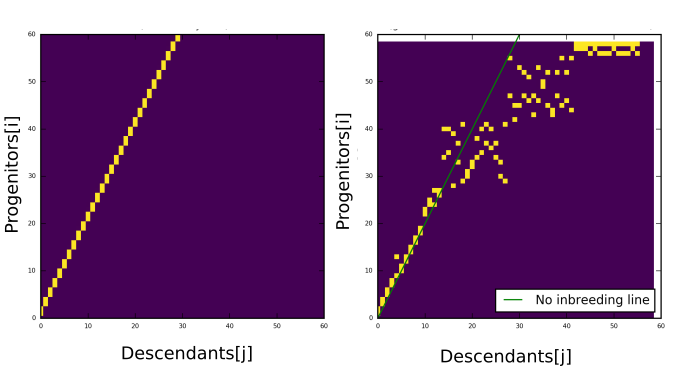}
\caption{{Left panel: Adjacency matrix of a full binary tree without inbreeding. Right panel: Adjacency matrix of the inbreeding tree realization in Figure \ref{fig_00} (bottom).}}
\label{fig_03}
\end{center}
\end{figure}

The case of a  binary tree is represented at the left of Figure \ref{fig_03}.  Let us compare it with what happens when we have an inbreeding tree, such as the one obtained in Figure \ref{fig_00} (bottom), which we show at the right side of Figure \ref{fig_03}. In the first case, {the lack of} inbreeding is reflected by the yellow line (which has slope 2), where none of the nodes share a progenitor. In the second case, it is interesting to observe that the yellow points separate from the slope corresponding to the di-graph, creating a unique distribution pattern {of each realization}. Adjacency matrix could be used as a useful measure of the degree of inbreeding with respect to the full binary tree.

\begin{figure}[htb!]
\begin{center}

\includegraphics[width=14cm]{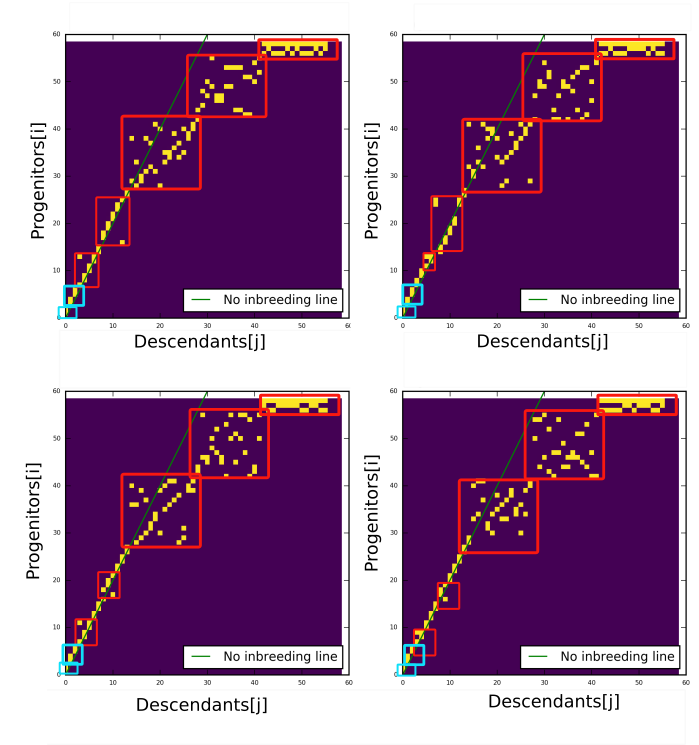}
\caption{Adjacency matrices corresponding to the trees presented in Figure \ref{fig_02}. In these plots, the distinctive imprint of each tree can be seen more clearly in the distributions of yellow dots on the dark background. The red rectangles indicate dots that must be in the same generation. The cyan rectangles, at the bottom of the matrices, belong to the first and second generations.}
\label{fig_07}
\end{center}
\end{figure}

The adjacency matrix acts as a fingerprint of each tree realization. For each graph, the plot of the adjacency matrix is unique. To illustrate this point, Figure \ref{fig_07} shows the adjacency matrices corresponding to the trees presented in Figure \ref{fig_02}. Each of them is different, and so are their corresponding graphs. In all the matrices, the number of ancestors at the beginning of the tree is low. As a result, a flattening of the curve appears as the number of the generation increases. The red rectangles indicate yellow points that must necessarily belong to the same generation. The cyan rectangles at the bottom of the graphs indicate the first and second generation from the trees, using the fact that we started from a binary tree.  These show that these graphs also indicate the number of generations from each tree: it is simply the number of rectangles. It can be seen more clearly from these plots that, even though the number of ancestors is the same, the edges in the tree are distributed differently. This is reflected in the different distribution of yellow points in the rectangles.

\begin{figure}[htb!]
\begin{center}
\includegraphics[width=14.5cm]{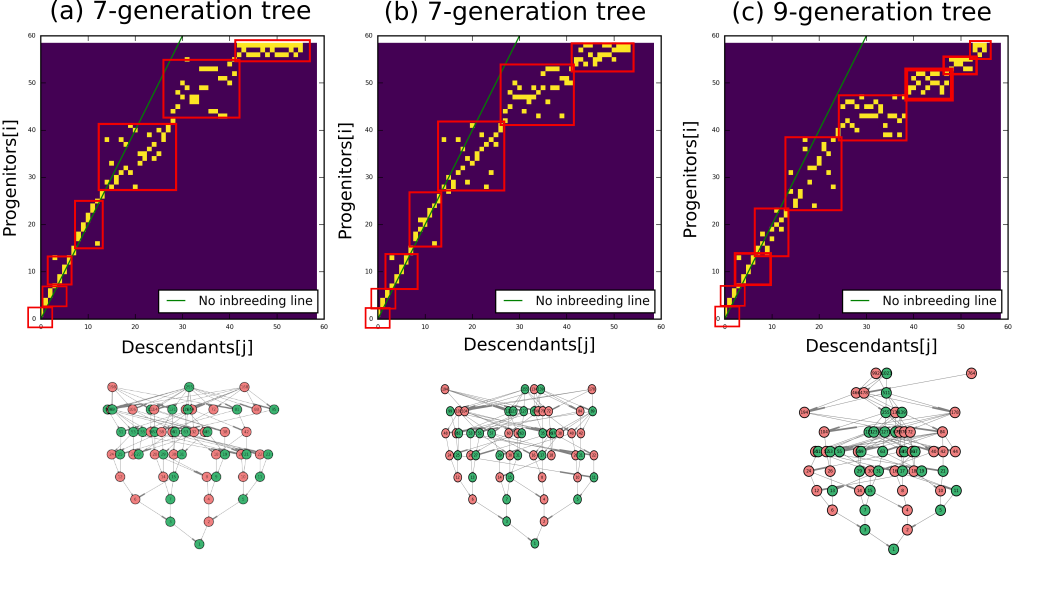}
\caption{Comparison of different adjacency matrices for three different trees (shown below each matrix) with the same number of nodes. Panels (a) and (b) correspond to 7-generation trees with the same number of nodes but different distribution in the generations. Panel (c) corresponds to a 9-generation tree. The red rectangles indicate yellow dots that must belong to the same generation. }
\label{fig_08}
\end{center}
\end{figure}

To further explore the information displayed in the adjacency matrices, we compare three cases: one of the trees that we have already presented, the other a 7-generation tree with the same total number of nodes but a different distribution, and then a 9-generation tree with the same number of nodes. The matrices and the trees are shown in Figure \ref{fig_08}. First, it can be clearly seen that the graph from panel (c) corresponds to a tree with more generations. Then, comparing panels (a) and (b) we can see that, for example, the tree associated with the matrix in panel (b) must have more nodes in the last generation. This further illustrates the usefulness of this graphic representation. 

{ Inspection of the figures shows that, clearly, the size of the rectangles is not uniform. The height of each rectangle is given by the number of ancestors present in the considered generation.  Within each rectangle, the uniformity of coverage indicates a high degree of inbreeding in that generation. For example, in all the plots from Figure \ref{fig_07}, the rectangle at the top, corresponding to the last generation, has a great degree of coverage, and, specifically, one of the lines is ``full'', implying that one node is linked to all the nodes in the following generation. }

\begin{figure}[htb!]
\begin{center}
\includegraphics[width=1.0\textwidth]{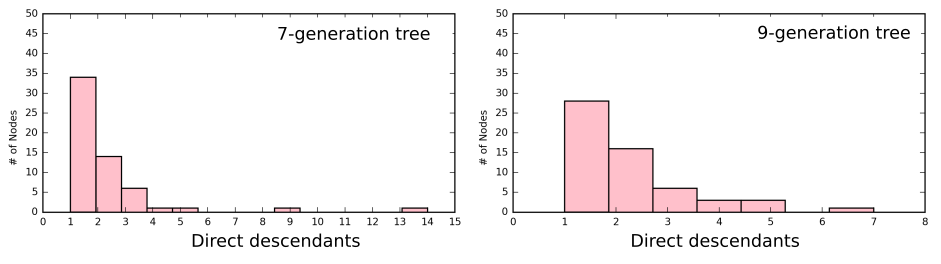}
\caption{Histograms with the number of descendants per node for two trees with the same number of nodes: (left) the 7-generation tree from Figure \ref{fig_02}  and (right) the 9-generation tree from Figure \ref{fig_08}.}
\label{fig_05}
\end{center}
\end{figure}

{Finally, we study other properties of the graph, such as the degree of connectivity of the nodes. By studying the distribution of the output degree of the nodes, we can observe also inbreeding reflected. This is shown in Figure \ref{fig_05}.} Here, we plot a histogram of the number of descendants associated with each node for two cases: on the left, the tree presented in Figure \ref{fig_00} (bottom), and on the right, for the 9-generation tree with the same number of nodes shown in Figure \ref{fig_08}. From the Figure, we can see that in the case of the 7-generation tree, the majority of the nodes have one descendant, some of them have two, and only a minority have more than 5. { We can see that there is one node with a significantly higher number of edges (14), which corresponds to the male node from the $n=7$ generation, the only male node at that stage, and thus connected to all the nodes from the $n=6$ generation.} For the 9-generation tree, the histogram is flatter and broader. This is associated with the fact that since this tree has the same number of nodes as a 7-generation tree, more edges are missing from a full binary tree, and therefore more nodes need to be linked to more than one element from the next generation.

\begin{figure}[htb!]
\begin{center}
\includegraphics[width=1.0\textwidth]{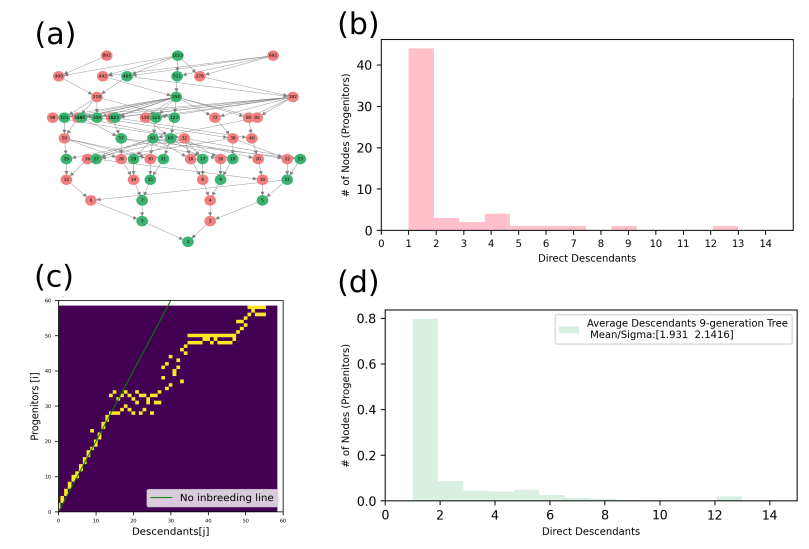}
\caption{{(a) One tree realization for a 9-generation tree with a different node distribution, its corresponding link histogram (b), and adjacency matrix (c). (d) Histogram of the output links  averaged over 50 tree realizations with the same node distribution as (a)}}
\label{fig_comp_9}
\end{center}
\end{figure}

{ We may also study global properties averaging different tree realizations, a feature that is included in the algorithm we present here. Fixing the number of ancestors to 58, as in the trees shown above, we can average the output of the degree of nodes for 50 different realizations of trees with the same number of nodes per generation, for trees of $5,6,7$ and $9$ generations. The realizations of the 7 and 9-generation trees have the same number of nodes in each generation as those shown in Figure \ref{fig_02} and \ref{fig_08}. We plot the resulting normalized histograms in Figure \ref{fig_average}, where we indicate the mean value and the standard deviation.

\begin{figure}[htb!]
\begin{center}
\includegraphics[width=1.0\textwidth]{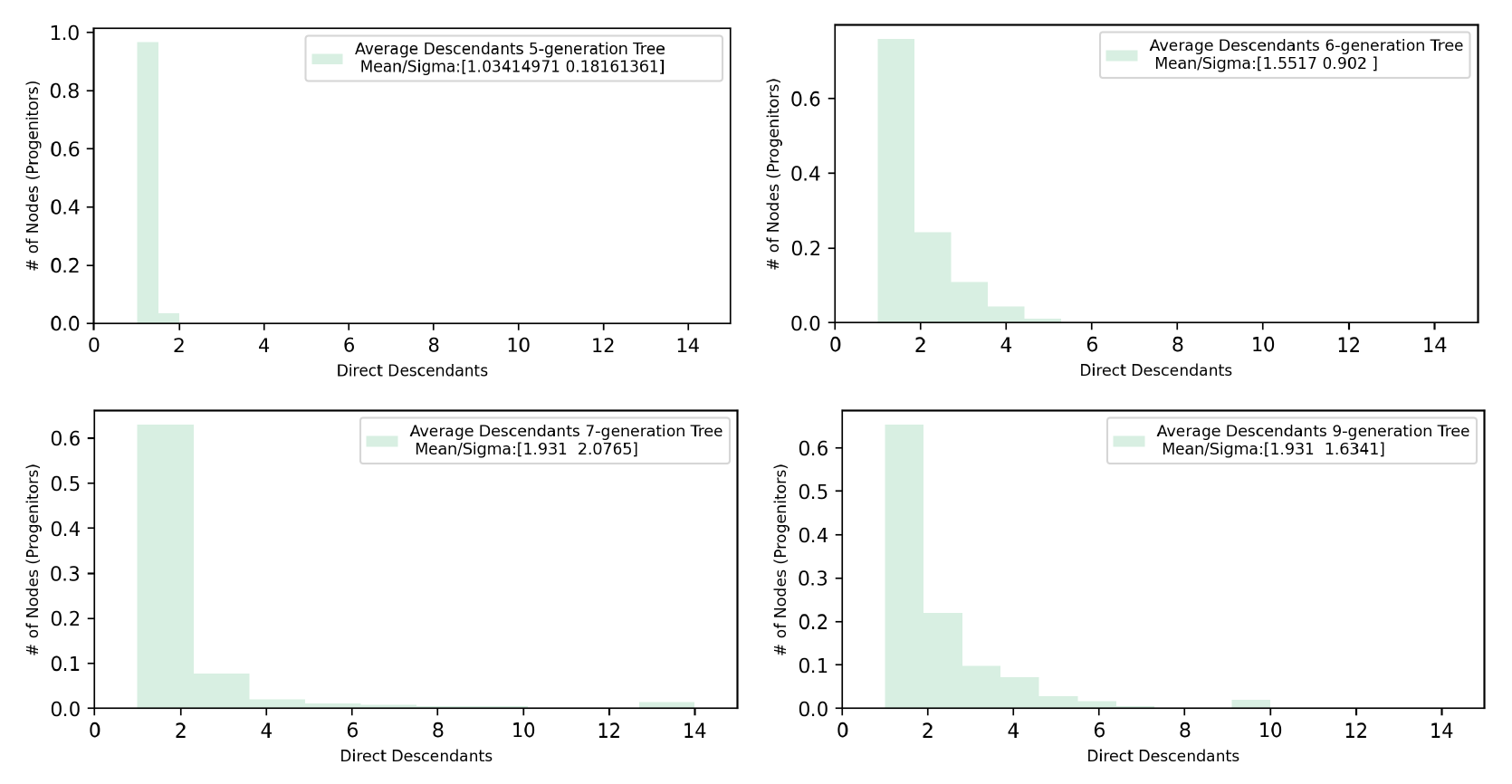}
\caption{{Histograms of the number of descendants per node averaged over 50 tree realizations with the same total number of ancestors (58) and a fixed number of nodes per generation, for 6,7,8 and 9 generations. The mean value and the standard deviation (``sigma'') are indicated. The 7-generation (9-generation) node distribution is the same as the tree realization from Figure \ref{fig_02}  (Figure \ref{fig_08} (c)).}}
\label{fig_average}
\end{center}
\end{figure}

Clearly, for the 5-generation trees, the histogram has a large peak at 1. This is simply due to the fact that not many nodes have been removed from the full binary 5-generation tree, which has 62 ancestors. As the proportion of removed nodes increases (in this case, as the same number of nodes is distributed in more generations), there is necessarily more inbreeding, which is reflected in a higher mean value,  and a higher standard deviation, a broader histogram. In these examples, the mean value of the 7 and 9-generation trees seems to be the same, but a quick inspection of the histograms shows that this is due to one node in the n=7-generation tree which has 14 links. If we do not take it into account, the mean value is lowered to 1.72 and the standard deviation is reduced to 1.33, lower than in the 9-generation case.  In general, we see that for a fixed value of generations and nodes, as the degree of inbreeding increases, the peak of the histogram decreases (i.e. there are more nodes with more than one descendant or, in other words, more descendants share direct ancestors.).

It is also worth noting that the histogram of the averages of the tree realizations depends not only on the number of nodes but on the distribution of these nodes in the generations. We illustrate this for the 9-generation case, where we compare the previously shown histogram with another average of realizations with 58 ancestors with different numbers of nodes per generation. In Figure \ref{fig_comp_9} we show an example of one such realization, its adjacency matrix, and node distribution histogram, and we also plot the histogram of the average over 50 realizations with the same node distribution. The average of output links (direct descendants) is practically the same, but the spread is different, in this case, because there are fewer nodes at the top of the tree, and there is an element with higher output links.  }

{ In this way, we have shown different features of this algorithm and how it allows for visualization and comparison of different types of trees. We have described how this algorithm makes it possible to obtain different tree representations and explore the connectivity information in each realization in different manners: either plotting the tree, studying the adjacency matrix or analyzing the connectivity between nodes. The algorithm also includes the possibility of averaging the connectivity among a set of different realizations, obtaining the corresponding histogram, and calculating the mean value and the standard deviation.}

\section{Conclusions}\label{conclu}

In the present work, we have presented an algorithm with a software implementation to build one possible ancestors' tree with its corresponding graph {(realization).
First, we defined three conditions for a tree of ancestors: (i) there is no overlapping or jumping between generations (ii) every node is linked to the previous and the next generation (iii) we assume a biparental reproductive scheme with two genders, and each node must have a progenitor of each gender. Then, we defined the steps for our algorithm: (1) we start from a full binary tree (2) we remove nodes (and links) from each generation (3) we complete the missing links, taking into account the conditions we have established, and obtain one realization of an inbreeding tree. There are many ways to do these steps. In the examples of trees of ancestors shown in this work, we required that the number of male and female nodes in each generation remained balanced, and we subtracted nodes in each generation following a markovian algorithm \cite{JARNE20191}. These conditions are the ones that we used in the algorithm, and may be changed by the user, to adapt to other cases, including different reproductive schemes, etc.}

Regarding the software implementation, this algorithm provides different tools to study the properties of the trees and their degree of inbreeding: it produces the tree itself, the adjacency matrix (which we have shown is a useful tool to visualize and distinguish trees), and a histogram illustrating the degree of connectivity of the nodes. We have shown how, from a simple ancestor distribution model, we can study { the different} corresponding graphs with different edges for the trees.  { We also include a function to average different tree realizations,  obtain an average link histogram, and calculate its mean value and standard deviation.} Moreover, the code gives the possibility of extending the algorithm and the mode and using these tools to perform different types of simulations and analysis. For example, our next step would be to add attributes to the nodes that could be considered as genes, or maybe diseases,  study the propagation of this feature along the tree, and compare it with the degree of inbreeding. 

{
  As for possible applications of present work, different kinds of trees (graphs) are used for different fields in science, such as decision trees for Machine Learning, Taxonomy trees in biology. Other fields involve, for example, Text Parsing Tree, Social Hierarchyor even Probability Trees \cite{Song2015, Hinchliff12764, COOPER2012161, Koski2015, FOX2002291}. The code presented here could be adapted and used, for instance, in interdisciplinary fields involving some of these areas. Given a set of data that may be mapped to a  tree of ancestors, the graphs shown here can be built and studied, as well as compared with variations which may also be constructed with the open code presented in this work.}

Finally, the code presented in this work is open source and easily adapted to study other situations relevant to current studies such as opinion spread models, disease propagation, etc. {In short, any problem where it is relevant to quantify all the different ways of originating a binary kinship structure.} We are thus certain that it will prove to be a useful tool for future studies.
 
\section*{Acknowledgements}

Present work was supported by CONICET and UNQ. F. A. G. A. is supported by SECyT UNLP and PICT 2018-02968.  This work was finished during the Covid-19 quarantine.

\appendix
\section{Code implementation} \label{code}
\Urlmuskip=0mu plus 1mu\relax
Code will be avaliable at the github repository upon publication: \\
$https://github.com/katejarne/Ancestors\_trees$

\Urlmuskip=0mu plus 1mu\relax
\bibliographystyle{elsarticle-num} 
\bibliography{mybibfile}







\end{document}